\documentclass[10pt,conference]{IEEEtran}
\usepackage{threeparttable,booktabs}
\usepackage{graphicx}
\usepackage{amsmath,amssymb,amsfonts}
\usepackage{algorithmic}
\usepackage{tabularx}
\usepackage{textcomp}
\usepackage{xcolor}
\usepackage{xspace}
\usepackage[acronym]{glossaries}
\usepackage{hyperref}
\usepackage[capitalise]{cleveref}
\usepackage{paralist}
\usepackage{cite}
\usepackage{listings}
\usepackage{tcolorbox} 
\usepackage{booktabs}
\usepackage{float}
\usepackage{tikz}
\usepackage{blindtext}
\usepackage{svg}

\newcommand{\TODO}[1]{\textcolor{red}{#1}\GenericWarning{}{LaTeX Warning: TODO: #1}}\newcommand\todo\TODO
\newcommand{\MODIFIED}[1]{\textcolor{blue}{#1}\GenericWarning{}{LaTeX Warning: MODIFIED: #1}}\newcommand\modified\MODIFIED

\def\BibTeX{{\rm B\kern-.05em{\sc i\kern-.025em b}\kern-.08em
    T\kern-.1667em\lower.7ex\hbox{E}\kern-.125emX}}
\begin{document}
\UseRawInputEncoding
\newcommand{\toolname}{\textsc{VeCoGen}\xspace}
\newcommand{\dataset}{\textsc{VeCoSet}\xspace}
\newcommand{\RQone}{How effective is \toolname in terms of generating formally verified C programs?}
\newcommand{\RQtwo}{How does providing a natural language or formal specification impact the effectiveness of \toolname? }
\newcommand{\RQthree}{What is the impact of changing parameters, prompts, and \glspl{llm} on the effectiveness of \toolname?}

\newacronym{llm}{LLM}{Large Language Model}
\newacronym{api}{API}{application programming interface}
\newacronym{cegis}{CEGIS}{CounterExample Guided Inductive Synthesis}
\newacronym{acsl}{ACSL}{ANSI/ISO C Specification Language}
\newacronym{apr}{APR}{Automatic Program Repair}
\newacronym{wp}{WP}{Weakest Precondition}
\newacronym{loc}{LoC}{Lines of Code}
\newacronym{vst}{VST}{Verified Software Toolchain}
\newacronym{gpt}{GPT}{Generative Pre-training Transformer}
\newacronym{ast}{AST}{abstract syntax tree}
\newacronym{sygus}{SyGuS}{Syntax-Guided Synthesis }
\newacronym{rte}{RTE}{Runtime Error}

\title{VeCoGen: Automating Generation of Formally Verified C Code with Large Language Models}

\author{\IEEEauthorblockN{Merlijn Sevenhuijsen}
\IEEEauthorblockA{\small \textit{Scania \& KTH Royal Institute of Technology} \\
S\"{o}dert\"{a}lje, Sweden \\
merlijn.sevenhuijsen@scania.com}
\and
\IEEEauthorblockN{Khashayar Etemadi}
\IEEEauthorblockA{\small \textit{KTH Royal Institute of Technology} \\
Stockholm, Sweden \\
khaes@kth.se}
\and
\IEEEauthorblockN{Mattias Nyberg}
\IEEEauthorblockA{\small \textit{Scania \& KTH Royal Institute of Technology} \\
S\"{o}dert\"{a}lje, Sweden \\
mattias.nyberg@scania.com}
}

\newcounter{inlineenum}
\renewcommand{\theinlineenum}{\alph{inlineenum}}
\newenvironment{inlineenum}
  {\unskip\ignorespaces\setcounter{inlineenum}{0}%
   \renewcommand{\item}{\refstepcounter{inlineenum}{\textit{\theinlineenum})~}}}
  {\ignorespacesafterend}
  
\definecolor{mGreen}{rgb}{0,0.6,0}
\definecolor{mGray}{rgb}{0.5,0.5,0.5}
\definecolor{mPurple}{rgb}{0.58,0,0.82}
\definecolor{backgroundColour}{rgb}{0.95,0.95,0.92}

\lstdefinestyle{CStyle}{
    backgroundcolor=\color{backgroundColour},   
    commentstyle=\color{mGreen},
    keywordstyle=\color{magenta},
    numberstyle=\tiny\color{mGray},
    stringstyle=\color{mPurple},
    basicstyle=\footnotesize,
    breakatwhitespace=false,         
    breaklines=true,                 
    captionpos=b,                    
    keepspaces=true,                 
    numbers=left,                    
    numbersep=5pt,                  
    showspaces=false,                
    showstringspaces=false,
    showtabs=false,                  
    tabsize=2,
    language=C,
}

\tcbset{
    conclusionbox/.style={
        colback=white,
        colframe=black,
        left=5pt,
        right=5pt,
        top=5pt,
        bottom=5pt,
        boxrule=0.5pt,
        arc=0pt,
    }
}

\maketitle
\begin{abstract}

Large language models have demonstrated impressive capabilities in generating code, yet they often produce programs with flaws or deviations from intended behavior, limiting their suitability for safety-critical applications. To address this limitation, this paper introduces \toolname, a novel tool that combines large language models with formal verification to automate the generation of formally verified C programs. \toolname takes a formal specification in ANSI/ISO C Specification Language, a natural language specification, and a set of test cases to attempt to generate a verified program. This program-generation process consists of two steps. First, \toolname generates an initial set of candidate programs. Secondly, the tool iteratively improves on previously generated candidates. If a candidate program meets the formal specification, then we are sure the program is correct. We evaluate \toolname on 15 problems presented in Codeforces competitions. On these problems, \toolname solves 13 problems. This work shows the potential of combining large language models with formal verification to automate program generation.
\end{abstract}


\section{Introduction}
\label{introduction}

\glspl{llm} have demonstrated versatility, excelling in various tasks \cite{brown_language_2020, chang_survey_2024, urlana_llms_2024, zeng_teaching_2024 }. One of the tasks where \glspl{llm} perform well is the generation of programs \cite{chen_evaluating_2021, liu_is_2023, lu_codexglue_2021}. Despite their impressive capabilities, \glspl{llm} often produce programs with errors or inconsistencies, making them unsuitable for applications requiring high assurance of correctness \cite{wang_is_2024}. This lack of trustworthiness poses a significant challenge to safety-critical domains where the correctness of programs is imperative. In the safety-critical domain, even minor software defects can have severe consequences, such as financial losses or threats to human life \cite{knight_safety_2002, klein_comprehensive_2014, leroy_formal_2009}. To address the lack of trustworthiness, the present paper introduces a new tool named \toolname, which combines \glspl{llm} with formal verification techniques to automatically generate C programs that are correct with respect to given specifications. 

\toolname is based upon a novel two-step process of initial code generation and iterative code improvement through feedback from a compiler and verifier. In the initial code generation step, \toolname generates an initial set of program candidates based on natural language specifications in English and formal specifications in \gls{acsl} \cite{baudin_acsl_2021}. The \gls{wp} and \gls{rte} plugins of Frama-C \cite{kirchner_frama-c_2015} then verify the correctness of the program candidates. If all generated program candidates fail compilation or verification, \toolname continues to the iterative code improvement step. In this step, \toolname parses the feedback from the compiler and verifier to guide the \gls{llm} in generating improved candidates. \toolname ensures that the generated program candidate is not only syntactically valid but also formally correct with respect to the formal specification. The tool can be downloaded from \href{https://github.com/ASSERT-KTH/Vecogen}{https://github.com/ASSERT-KTH/Vecogen}. 

Traditional works, not utilizing \glspl{llm}, have addressed the challenge of generating programs automatically \cite{jha_theory_2016}, but they often face scalability issues \cite{srivastava_program_2010, solar-lezama_program_2008}. \glspl{llm} offer a promising solution to the scalability issues of generating a program that meets the specifications, as explored in prior research. Mukherjee and Delaware \cite{mukherjee_towards_2024} employ \glspl{llm} along with human intervention to synthesize and verify C programs, demonstrating capability in handling complex scenarios. 
Similarly, Patil et al. \cite{patil_towards_2024} propose \emph{spec2code}, a framework that combines \glspl{llm} with critics to iteratively synthesize programs. However, these existing approaches either rely on manual feedback to the \gls{llm} or do not have a tool that implements the fully automatic verified code generation. \toolname is the first \gls{llm}-based tool that fully automatically generates and verifies C code. 

We evaluate \toolname on 15 competitive programming problems to assess its effectiveness in generating formally verified C programs. \toolname solves 13 out of 15 problems, demonstrating its ability to generate formally verified code. This initial benchmarking showcases the potential of generating formally verified C code automatically using \toolname, allowing for use in safety-critical software development.

The paper contains the following contributions:
\begin{enumerate}
    \item \toolname, a novel \gls{llm}-based code generation tool for iteratively generating formally verified C code. 
    \item The evaluation of \toolname on a collection of 15 competitive programming problems. 
    \item An analysis of the impact of changing the configuration of the tool, including the specification type, the number of generated programs per \gls{llm}-invocation, the temperature, the zero- or one-shot prompting, and the choice of \gls{llm}. 
\end{enumerate}

The rest of this paper is organized as follows. \Cref{sec:background} provides background on formal verification and \gls{llm}-based code generation. \Cref{sec:approach} describes the design and implementation of \toolname. \Cref{sec:experimentalmethodology} outlines the experimental methodology, and \Cref{sec:results} presents the results. \Cref{sec:threats} discusses threats to validity, \Cref{sec:related-work} presents related work, and \Cref{sec:conclusion} concludes the paper.

\section{Background}
\label{sec:background}

\subsection{ANSI/ISO C Specification Language}
\label{background:ACSL}

\begin{figure}[t]
    \centering
    \begin{lstlisting}[language=C, style=CStyle, basicstyle=\scriptsize\ttfamily, breaklines=true]
/*@
    requires 1000 >= x >= 0 && 1000 >= y >= 0;
    requires \valid(result);
    assigns *result;
    ensures *result == x + y;
*/
void add_positive(int x, int y, int* result);
\end{lstlisting}
    \caption{Formal specification for an ``add\_positive'' program.}
    \label{fig:formal-spec}
    \vspace{-2mm}
\end{figure}

ANSI/ISO C Specification Language (ACSL) is a formal specification language for describing the desired behavior of a function in C \cite{baudin_acsl_2021}. \Cref{fig:formal-spec} presents an example of an \gls{acsl} specification for a program that computes the sum of two positive integers. The formal specification for this ``add\_positive'' function uses three clause types: \texttt{requires}, \texttt{assigns}, and \texttt{ensures}. 

The \texttt{requires} clause specifies preconditions that must be met before running the function. In the  ``add\_positive'' example, the first precondition states that both input values, \texttt{x} and \texttt{y}, must be positive integers (line 2). Additionally, the second precondition states that the result variable must point to a valid memory location (line 3). The \texttt{assigns} clause defines which memory locations the function can modify while executing. The function is permitted to only modify the memory location pointed to by the \texttt{result} variable (line 4). The \texttt{ensures} clause defines postconditions that must hold after the function completes execution. The postcondition specifies that the output variable, \texttt{result}, must be equal to the sum of \texttt{x} and \texttt{y} (line 5).

A program that verifies against this specification is guaranteed to implement the intended behavior correctly. In addition to the \texttt{requires}, \texttt{assigns}, and \texttt{ensures} clauses, \gls{acsl} supports many other types. These extra clauses are explained in detail in the official \gls{acsl} documentation\footnote{For further details on ACSL clauses, refer to the official documentation at \url{https://frama-c.com/html/acsl.html}.}.

\subsection{Frama-C}
\label{background:Frama-C}

\texttt{Frama-C} is a platform for the static analysis and formal verification of C programs \cite{cuoq_frama-c_2012, kirchner_frama-c_2015}, used in safety-critical projects \cite{dordowsky_experimental_2015, brito_program_2010, ung_post-hoc_2024}. The \gls{wp} plugin of Frama-C, inspired by the principles of Hoare Logic \cite{hoare_axiomatic_1969}, verifies functional properties by generating proof obligations based on \gls{acsl} specifications. These obligations are then translated into logical goals using the \texttt{Why} platform \cite{filliatre_why_2003}. The \gls{rte} plugin complements the \gls{wp} plugin by automatically generating goals to check for runtime errors, such as integer overflows. 

Automated theorem provers like \texttt{Alt-Ergo}, \texttt{CVC}, and \texttt{Z3} \cite{de_moura_z3_2008} attempt to prove the logical goals. The theorem provers validate these goals within specified limits. If a goal cannot be proven within the given timeout and computational limits, the plugin provides output to help developers refine their code or specifications. Programs are considered formally verified with respect to the formal specification once all goals generated by the \gls{wp} and \gls{rte} plugins are successfully proven.

\subsection{Large Language Models}
\label{background:LLM}

Large Language Models (LLMs) are machine learning models with a large number of parameters, trained on a vast corpus of data. State-of-the-art \glspl{llm}, such as GPT-4o, are built on the Transformer architecture \cite{vaswani_attention_2017} and use decoder-only models \cite{vaswani_attention_2017}. The decoder-only models generate text by continuously predicting the next token, consisting of a small set of characters. After generating a token, the generated token is appended to the input. The \gls{llm} continues with iteratively producing more tokens until a stop token is predicted as the next token. These decoder-only models have shown highly promising results on code-related tasks \cite{chen_evaluating_2021}, and we employ them in this study for generating formally verified C code.

\section{VeCoGen Approach}
\label{sec:approach}

\begin{figure*}[!t]
    \centering
    \includegraphics[width=\textwidth]{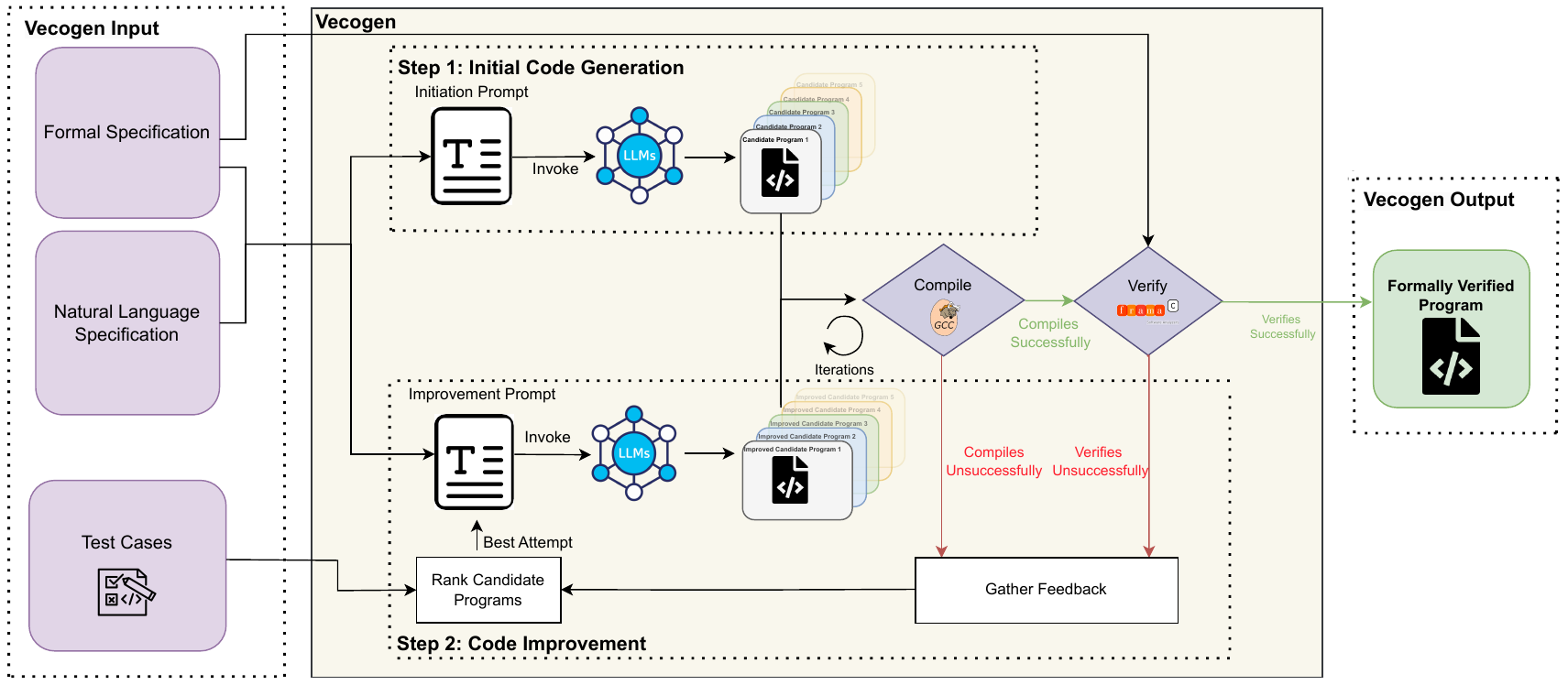}
    \caption{Outline of \toolname: \toolname uses formal and natural language specifications to generate a program that meets the formal specification.}
    \label{fig:vecogen-tool}
\vspace{-4mm}
\end{figure*}

This section describes \toolname, an iterative \gls{llm}-based tool aimed at automatically generating formally verified C code. \toolname uses specifications and test cases to generate a C program that meets the specifications. This approach guarantees the correctness of \gls{llm}-generated code, which is a major goal in \gls{llm}-based code generation literature \cite{spiess_calibration_2024,tihanyi_formai_2023,liu_exploring_2024}.

\subsection{VeCoGen Overview}
\label{approach-overview}

\Cref{fig:vecogen-tool} presents an overview of how \toolname works. The code generation process using \toolname consists of four important parts:
\begin{inparaenum}[(1)] 
    \item inputs, which are the program specifications, see \Cref{approach-problemspecification};
    \item initial code generation using \toolname, see \Cref{approach-initialcodegeneration};
    \item code improvement using \toolname,  see \Cref{approach-codeimprovement};
    \item output, which is a formally verified program, see \Cref{approach-output}.
\end{inparaenum}

The tool, \toolname, performs two steps: \emph{initial code generation} and \emph{code improvement}. The initial code generation step generates a set of initial programs by invoking an \gls{llm} through an \emph{initialization prompt}. If none of the programs generated at this step meet the given formal specification, then \toolname continues onto step two. Within the code improvement step, \toolname iteratively improves the best previously generated program by invoking an \gls{llm} using an \emph{``improvement prompt''}. This prompt asks the \gls{llm} to improve a candidate program using feedback from a compiler and verifier. The tool iteratively prompts an \gls{llm} using previously generated feedback until the \gls{llm} generates a program that satisfies the formal specification. \toolname is the first \gls{llm}-based tool that automatically generates formally verified C code. 


\subsection{VeCoGen Input}
\label{approach-problemspecification}

The search for a formally verified program consists of three inputs:
\begin{inparaenum}[(1)] 
    \item a formal specification given in \gls{acsl},
    \item a natural language specification in English, and
    \item a set of unit test cases in C. 
\end{inparaenum} The first input is a formal specification used to verify candidate programs. The goal of the tool is to generate a program that meets the \gls{acsl} specification. 

\begin{figure}[t]
    \centering
    \begin{lstlisting}[basicstyle=\scriptsize\ttfamily, breaklines=true, escapeinside={(*@}{@*)}]
Write a function to compute the sum of two positive integers  and store the result at a specified memory location.
Input:
    Two positive integers `x` and `y` ( (*@ $ 1 \leq x, y \leq  10^6 $ @*) ) and a pointer `result` to store the sum.
Output:
    The function writes the sum of `x` and `y` to the memory location pointed to by `result`.
    \end{lstlisting}
    \caption{Natural language specification for the add\_positive program.}
    \label{fig:nl-spec}
    \vspace{-3mm}
\end{figure}

The second input is a natural language specification, which is an informal description of the desired behavior of a program. This type of specification conveys the functionality and purpose of the code to the \gls{llm} in natural language. \Cref{fig:nl-spec} shows the natural language description for the ``add\_positive'' program defined in \Cref{background:ACSL}. It describes the intended behavior of the program as well as its input and output.

Besides the formal and natural language specifications, we also define a function signature to specify the interface of the program. The signature defines the function name and the input and output parameters. We append this function signature to the specifications.

Lastly, \toolname requires a set of unit test cases. Each test case specifies an input along with the expected output. If the output of a program aligns with the expected output, then the test passes. If the output differs, then the unit test fails. \toolname relies on these test cases during the iterative code improvement phase to progressively refine the generated programs (see \Cref{approach-codeimprovement} for more details). The test cases, natural language, formal specification, and function signature must be consistent. 

\subsection{Step 1: Initial Code Generation}
\label{approach-initialcodegeneration}

\begin{figure}[t]
    \centering
    \includegraphics[width=0.5\textwidth]{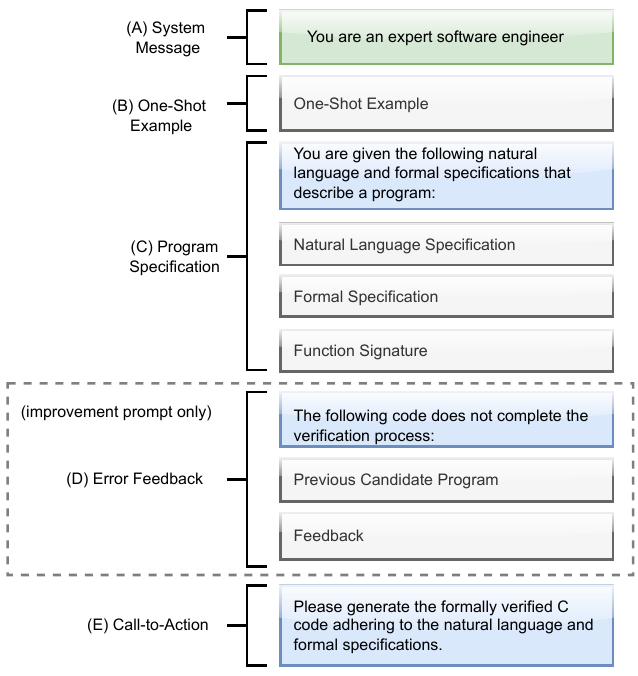}
    \caption{Prompt structure used in \toolname for generating formally verified C programs.}
    \label{fig:initiationprompt}
\end{figure}

Based on the two specification types, \toolname crafts a prompt to invoke an \gls{llm}. \toolname uses two types of prompts: the \emph{initialization prompt} and the \emph{improvement prompt}. \Cref{fig:initiationprompt} presents the outline of both prompt types. Both types contain a system message, a one-shot example, the specifications, and a call to action. The improvement prompt also includes error feedback on a previous attempt (see \Cref{approach-codeimprovement} for details). The parts of the prompt highlighted in green and blue remain the same throughout each prompt. Green indicates the role of the \gls{llm}, and blue indicates unchanged text explaining the task to the \gls{llm}. The gray parts indicate problem-specific information.




In the initial code generation step, \toolname uses the initialization prompt. This prompt type begins with a system message (A) that assigns an expert software engineer role to the \gls{llm}. Next, we include a one-shot example (B). The one-shot example provides one single detailed instance of the desired input-output of the \gls{llm}. We use the one-shot example to leverage in-context learning, as this explains the task and expected format to the \gls{llm} \cite{brown_language_2020}. This manually created one-shot example includes a natural language specification, a formal specification, a function signature, and a correct implementation. The one-shot example\footnote{The one-shot example can be seen at \url{https://github.com/ASSERT-KTH/Vecogen/blob/main/prompts/one_shot_example.txt}.} remains unchanged across all problems.

The third part of the prompt presents the specifications of the desired program (C). These program specifications include a natural language specification, an \gls{acsl} specification, a function signature, and a text explaining to the \gls{llm} what the model has to do with these. The function signature presents the input and output format of the desired program. The third part of the prompt outlines the specifications for the desired program (C). These specifications include four components: a description in natural language, an \gls{acsl} specification, a function signature, and instructions for the \gls{llm}. Within the instruction, the prompt specifies what to do with the specifications: to take the provided specifications and generate a program that adheres to them. Additionally, it includes constraints such as avoiding loops in the generated program. 

Part (D) of the prompt is not included in the initialization prompt as we have no previously generated candidate programs in this step. Lastly, both prompt types include a \emph{Call to Action} (E) instructing the model to generate a formally verified C program that meets the provided specification.

Using the initialization prompt, \toolname invokes the \gls{llm} to generate candidate programs. We check the correctness of each candidate in two steps. First, we compile the candidate. If the compilation is successful, \toolname employs the \gls{wp} and \gls{rte} plugins of Frama-C to prove that the candidate meets the formal specification. If any candidate passes these checks, it is considered a correct program, and the code generation is successfully completed. Otherwise, the candidates are sent to the next step for an iterative improvement until a correct program is generated.

\subsection{Step 2: Code Improvement}
\label{approach-codeimprovement}
In the code improvement step, incorrect candidates are iteratively repaired to synthesize a formally verified program. As seen in \Cref{fig:initiationprompt}, the improvement prompt has the same system message, one-shot example, and specification as the initialization prompt. In addition to the initialization prompt, the improvement prompt contains a previously generated candidate program along with associated compilation or verification feedback. This feedback provides the \gls{llm} with valuable information to enhance its previously generated attempt.

The improvement prompt only contains one incorrect program. As multiple candidates are generated at each iteration, we have to pick one of the candidates to include in the next improvement prompt. Selecting a single candidate ensures that the \gls{llm} is not overwhelmed with conflicting information from multiple programs. \toolname randomly selects one of the programs that pass the highest percentage of unit test cases. We choose the most promising candidate to increase the likelihood of quickly converging to a verified program. The \gls{llm} then generates improved versions of this program. These revised programs are then compiled and verified. The iterative process of generating new programs based on previous incorrect candidates is repeated until a formally verified program is found or a maximum of ten iterations is reached.

\subsection{\toolname Output}
\label{approach-output}

The output of \toolname consists of a formally verified C program that adheres to the provided formal specification or, if unsuccessful, the last generated candidate program. Additionally, \toolname produces a detailed log file for each run. If \toolname is successful, then the generated program is guaranteed to be both syntactically correct and semantically correct with respect to the formal specification.

\subsection{Implementation}
\label{approach-implementation}

\toolname is written in Python and uses GCC for compilation and the Frama-C plugins \gls{wp} and \gls{rte} for verification. Solvers attempt to prove the goals generated by the two Frama-C plugins. \toolname has been tested with solvers Alt-Ergo, CVC4, and Z3. Frama-C was chosen for its strong verification capabilities, especially in safety-critical systems. Due to the absence of complex control structures like loops, Frama-C offers automatic verification with theorem provers. If more complexity is introduced, then Frama-C would require help to prove that a program adheres to a formal specification.

\toolname is able to employ any \gls{llm} with an \gls{api}. Currently, \toolname is configured to use the following \glspl{llm}: GPT-3.5-turbo, GPT-4o, Llama-3.1-70B. The default \gls{llm} in \toolname is GPT-3.5-turbo, selected for its balance of cost-effectiveness and suitable performance for generating C programs. 

\toolname generates ten programs per \gls{llm} invocation. We pick ten, as our experiments show that generating ten programs per request balances diversity in generated candidate programs and the number of duplicates between the generated candidates. To promote the diversity of the programs, \toolname uses a sampling \emph{temperature} of 1. The temperature controls the randomness of an \gls{llm}  when generating output by adjusting the probability distribution of predicted tokens. The values for the temperature range from 0 to 1, where lower values make the output more deterministic, and higher values also introduce greater diversity by predicting less likely tokens. This encourages the \gls{llm} to explore varied solutions and minimizes duplicates. During the code improvement step, \toolname uses at most ten iterations. We use ten iterations to give the \gls{llm} multiple chances to refine incorrect programs based on feedback from the compiler and verifier. 

\section{Experimental Methodology}
 \label{sec:experimentalmethodology} 
 To assess the effectiveness of \toolname and the impact of specification types on its performance, we design an experimental methodology guided by research questions to evaluate its ability to generate formally verified C.

\subsection{Research Questions}
We define three research questions to evaluate the effectiveness of \toolname in generating formally verified C programs:
\begin{itemize}
    \item [\textbf{RQ1 (effectiveness)}:] \RQone \xspace As a metric, we count the number of specifications for which \toolname generates a formally verified program. Additionally, we investigate the number of problems solved after initial code generation, the number of solved problems after different numbers of code improvement iterations, and the total time taken. If a solution is generated, we also present metrics of the solution.
    \item [\textbf{RQ2 (specification type impact)}:] \RQtwo \xspace We analyze the number of successfully generated verified programs based on different specification inputs. Specifically, we investigate using only natural language specification, only formal specification, and both specification types.
    \item [\textbf{RQ3 (ablation study):}] \RQthree \xspace Specifically, we study the impact of employing different search strategies in terms of number of candidates, temperature, iterations, the use one-shotting, and using different \glspl{llm}.
\end{itemize}
\subsection{Study Subjects}
To evaluate the performance of \toolname, we create a dataset called \dataset. \dataset consists of formal specifications, natural language specifications, and test cases for 15 problems selected from Codeforces online competitions \cite{noauthor_codeforces_nodate}. \textsc{Code4Bench}\cite{amirabbas_code4bench_2019} bundles problems used in Codeforces, from which we select problems based on the following criteria:
\begin{enumerate}
    \item The problem has a ground-truth accepted solution in the C language. This ensures that we have a reference solution.    
    \item The ground-truth solution to the problem is limited to a single function that does not use loops.
    \item A formal verification expert must be able to manually specify the problem in \gls{acsl} within three hours of work to keep the study manageable, as explained below.
\end{enumerate}

Furthermore, we exclude problems involving loops due to two reasons: 
\begin{inparaenum}[(1)] 
    \item Including problems that contain a solution with loops often requires additional loop invariants to formally verify the file. Writing these loop invariants typically requires manual effort and domain knowledge. Therefore, we focus our study on programs without loops to make our code generation approach automatic.
    \item In safety-critical domains like automotive, loops are often restricted due to predictability and reliability concerns.  Standards like \texttt{MISRA-C} \cite{MISRA2004} impose strict limits on loop usage to prevent issues such as unbounded execution and timing unpredictability. Excluding loops from the analysis aligns with these safety practices, supporting our goal of automatic code generation for such domains.
\end{inparaenum}

From the resulting set of 77 problems, we randomly pick 15 to include in \dataset. Problems from Code4Bench include natural language descriptions and test cases, which we use directly in \dataset without modification. However, formal specifications are not included in Code4Bench. Therefore, we manually create the formal specifications and check their correctness and completeness using two methods. First, we validate that the ground-truth solution meets the formal specification. Second, another formal verification expert confirms the consistency and completeness of each specification. We verify that the full program behavior is captured in the formal specification to ensure that only programs that capture the intended behavior are formally verified.

\begin{figure}[t]
    \centering
    \begin{lstlisting}[language=C, escapeinside={(*@}{@*)}, style=CStyle, basicstyle=\scriptsize\ttfamily, breaklines=true]
(*@\textcolor{blue}{Original code in main function}@*)
#include<stdio.h>  
int main()  
{  
    long long n;  
    scanf("%I64d", &n);  
    long long result;  
    result = n / 2;  
    if(n % 2 == 0)  
        printf("%I64d", result);  
    else  
        printf("%I64d", result + 1);  
}
(*@    \hrulefill@*)
(*@\textcolor{blue}{Transformed code as a function}@*)
void calculateMinimumBrainsForStrategy(long long N, long long *out)
{
    if (N % 2 == 0)
        *out = (N / 2);
    else
        *out = (N / 2) + 1;
}
\end{lstlisting}
\caption{A solution to a Codeforces problem in C, before and after transforming it to a function.}
    \label{fig:solution-combined}
\end{figure}

Solutions in Codeforces make use of the standard input and output. We transform the ground-truth solution into a single-function program to facilitate formal verification. This allows us to write formal specifications for each transformed function. Additionally, using functions is common practice in real-world projects. For example, \Cref{fig:solution-combined} presents a solution taken from Code4bench and its transformed functional version.

In the manual transformation procedure, we perform five steps: 
\begin{inparaenum}[(1)] 
    \item We create a function signature. This is a void function with a relevant name. 
    \item We introduce a parameter in the function signature for each value read from the standard input. This allows the function to receive input values as arguments rather than through \texttt{scanf} statements.
    \item We introduce a parameter in the function signature for each value printed to the standard output. For this, we utilize an output pointer to capture results.
    \item We replace any \texttt{printf} and \texttt{scanf} statements and replace them with assignments to the designated input and output parameters.
    \item We remove the include \texttt{\#include<stdio.h>} statement, as standard I/O functions are no longer used in the transformed function.
\end{inparaenum}

\begin{table}[t]
\caption{Summary of the problems present in \dataset.}

\begin{center}
\begin{tabular}{l|ccc}
    \toprule
    & \textbf{Median} & \textbf{Min} & \textbf{Max} \\
    \midrule
    Natural Language Spec. Size (Tokens) & 249 & 122 & 532 \\
    Formal Spec. Size (LoC) & 25 & 11 & 50 \\
    Formal Spec. Size (Clauses) & 7 & 4 & 14 \\
    Ground-truth Solution Size (LoC) & 9 & 1 & 20 \\
    Number of Test Cases & 34 & 5 & 104 \\
    \bottomrule
\end{tabular}
\label{tab:dataset-stats}
\end{center}
\end{table}

\Cref{tab:dataset-stats} summarizes features of the 15 problems in our dataset. It shows the size of the natural language specification, the size of the formal specification, the length of the ground-truth solution, and the number of test cases. The table presents the median value and the minimal and maximal values for each of these metrics. This information underscores the variety of problems present in \dataset.

The natural language specification size is measured in tokens. For each natural language description, we use the GPT-3.5 tokenizer to count the number of tokens. For example, the natural language specification presented in \Cref{fig:initiationprompt} consists of 141 tokens. The number of clauses in the formal specifications varies from 4 to 14 in \dataset, which shows the diversity in the level of detail needed to capture their requirements accurately. Problems in Codeforces are rated by difficulty to match user skill. The problems in \dataset contain problems classified as easy to medium.

\subsection{Protocol for RQ1 (effectiveness)}
To answer RQ1, we use \toolname to generate verified C programs for all 15 problems in \dataset. If \toolname generates a C program that meets the formal specification, then we consider the problem solved. We define the effectiveness of \toolname as the number and ratio of problems for which our tool solves the problem. This metric allows us to evaluate the effectiveness of \toolname to generate verified C programs.

We consider the number of problems solved after initial code generation as a baseline. Then, we track the number of solved problems after each code improvement iteration, up to a maximum of ten iterations or until all problems are solved. Additionally, we count metrics of the solution, such as the number of lines of code and the verification time required. Lastly, we capture the total runtime for attempting to generate and verify candidate solutions. These metrics provide insights into the performance and potential integration of \toolname. For unsolved problems, we manually analyze the generated code to identify the cause of failure.

\subsection{Protocol for RQ2 (specification type impact)}
To answer RQ2, we evaluate the impact of different specification types on the effectiveness of \toolname in generating verified C programs. For each of the 15 problems in \dataset, we run \toolname using three different specification types as inputs: 
\begin{inparaenum}[(1)] 
    \item only the natural language specification
    \item only the formal specification, and 
    \item both natural language and formal specifications combined. 
\end{inparaenum}

When only natural language specifications are used to generate code programs, the generated programs are attempted to be verified with respect to the formal specification. If this fails, then only natural language feedback is given back to the \gls{llm}. This feedback states that the code does not verify and that the \gls{llm} must improve on the code.


When running \toolname with each specification type, we record the number of problems solved after the initial code generation step and after iterative improvements. Additionally, we measure the total runtime used when generating programs using \toolname. Using these statistics, we evaluate the effectiveness and time efficiency using different specification types as input to \toolname.

\subsection{Protocol for RQ3 (ablation study)}
To answer RQ3, we perform an ablation study to evaluate how different configurations, such as including a one-shot example and using different \glspl{llm}, affect the performance of \toolname. We vary the number of candidate programs generated per \gls{llm} invocation, the used temperature of the \gls{llm} when generating candidates, whether or not a one-shot example is included, and the specific \gls{llm} employed.

The first (and default) configuration acts as a baseline, consisting of ten candidate programs per invocation, ten code improvement iterations, a temperature of 1, and a one-shot example. We create the second configuration to investigate the effect of taking the most promising candidate at each iteration. This configuration uses a temperature of 0 to get the best possible candidate at each iteration. This leads to little variety between generated program candidates, so we only generate one program candidate per invocation.

The third configuration investigates the effect of generating many candidate programs without using code improvement iterations. It generates 100 candidate programs per invocation, a commonly used number in the literature \cite{chen_evaluating_2021, stechly_gpt-4_2023}.

The fourth configuration explores providing a one-shot example, examining the impact of giving an example of the task and output format on the effectiveness of \toolname.

Lastly, we define configurations to use \toolname with different \glspl{llm}. This assesses the impact of using open-source (Llama-3.1-70B) and closed-source (GPT-3.5-turbo, and GPT-4o) \glspl{llm}. This provides insights into how the choice of \gls{llm} influences the formally verified program generation process.

For each configuration, we count the total number of problems solved by \toolname after completing the initial code generation and iterative improvement steps. By comparing the results of the defined configurations, we analyze the effect of changing information included in the prompt. Additionally, we examine the time each configuration needs to generate a verifying program to assess the impact on the performance and efficiency of \toolname.

\section{Experimental Results}
\label{sec:results}

\subsection{Results for RQ1}
\label{sec:resultsrq1}

\begin{table*}[!t]
\centering
\scriptsize
\caption{Effectiveness of \toolname on different problems of \dataset.}
\begin{tabular}{rl|rrrr}
    \toprule
    \textbf{Problem\_ID} & \textbf{Problem\_Name} &  \textbf{SS\_Iter} & \textbf{SS\_LoC} & \textbf{SS\_Ver\_Time\tnote{1}} & \textbf{Tot\_Time\tnote{1}} \\
    \midrule
    \href{https://codeforces.com/problemset/problem/581/A}{1}  & Vasya the Hipster & 0  & 2  & 2.2 s & 11 s \\
    \href{https://codeforces.com/problemset/problem/617/A}{2}  & Elephant & 0  & 5  & 1.6 s & 6 s \\
    \href{https://codeforces.com/problemset/problem/630/A}{3}  & Again Twenty Five! & 0  & 1  & 0.8 s & 6 s \\
    \href{https://codeforces.com/problemset/problem/638/A}{4}  & Home Numbers & 0  & 5  & 2.1 s & 129 s \\
    \href{https://codeforces.com/problemset/problem/690/A1}{5}  & Collective Mindsets & 0  & 1  & 1.2 s & 5 s \\
    \href{https://codeforces.com/problemset/problem/723/A}{6}  & The New Year: Meeting Friends & 3  & 14  & 4.2 s & 388 s \\
    \href{https://codeforces.com/problemset/problem/742/A}{7}  & Arpa’s hard exam & 0  & 14  & 1.3 s & 25 s \\
    \href{https://codeforces.com/problemset/problem/746/A}{8}  & Compote & 0  & 6  & 2.0 s & 103 s \\
    \href{https://codeforces.com/problemset/problem/760/A}{9}  & Petr and a calendar & 1  & 5  & 2.5 s & 213 s \\
    \href{https://codeforces.com/problemset/problem/151/A}{10} & Soft Drinking & 0  & 10  & 15.1 s & 35 s \\
    \href{https://codeforces.com/problemset/problem/168/A}{11} & Wizards and Demonstration & 0  & 8  & 3.5 s & 40 s \\
    \href{https://codeforces.com/problemset/problem/194/A}{12} & Exams & -- & --  & -- & 1,256 s \\
    \href{https://codeforces.com/problemset/problem/199/A}{13} & Hexadecimal's theorem & 1  & 3  & 1.2 s & 129 s \\
    \href{https://codeforces.com/problemset/problem/228/a}{14} & Is your horseshoe on the other hoof? & 1  & 2  & 2.1 s & 286 s \\
    \href{https://codeforces.com/problemset/problem/259/b}{15} & Little Elephant and Magic Square & -- & --  & -- & 1,090 s \\
    \bottomrule
\end{tabular}
\label{fig:effectiveness_table_rq1}
\vspace{-5mm}

\end{table*}

\begin{table}
\caption{Effectiveness of \toolname using different specification types.}
\centering

\begin{tabular}{l|rr|r}
    \toprule
    \textbf{Specification\_Type} & \textbf{Initial} & \textbf{Improvement} & \textbf{Tot\_Time} \\
    \midrule
    Both (default) & 9 & 13 & 3,722 s \\
    Natural Language & 10 & 13 & 5,447 s \\
    Formal & 7 & 8 & 10,538 s \\
    \bottomrule
\end{tabular}
\label{tab:specification_types_comparison}
\vspace{-5mm}
\end{table}

\Cref{fig:effectiveness_table_rq1} presents information about the problems and results\footnote{The results to all experiments conducted can be found at \url{https://github.com/ASSERT-KTH/Vecogen/tree/vecogen-results/results}.} using \toolname. Specifically, it presents generated solutions using the default configuration of \toolname. Column ``Problem\_ID'' mentions the problem identifier, and column ``Problem\_Name'' presents the name of the problem as presented in Codeforces. 
Column ``SS\_Iter'' presents the number of code improvement iterations that \toolname uses to synthesize a verifying solution. If 0, then the problem is solved in the initial code generation step. If \toolname does not solve the problem, we indicate this using ``--''. For the solved problems, column ``SS\_LoC''  shows the \gls{loc} of the synthesized solution. Column ``SS\_Ver\_Time'' presents the time needed to formally verify the synthesized solution. Lastly, column ``Tot\_Time'' presents the total time spent to synthesize a solution to the problem. The total time includes the total runtime of \toolname, Frama-C, and the \gls{llm} combined.

Column ``SS\_Iter'' shows the tool solves 9 of the 15 problems in \dataset during initial code generation. After the iterative code improvement step, the number of solved problems increases to 13. Notably, three problems (9, 13, and 14) are solved after the first code improvement iteration, and problem six is solved after the third iteration. This shows the effectiveness of the iterative approach of \toolname, as many problems are solved during improvement attempts.

When inspecting the \gls{loc} of the synthesized solution in the column ``SS\_LoC'', we see that \toolname generates solutions of varying lengths. The shortest solution generated is one line long, while the longest synthesized solution contains 14 \gls{loc}. Similarly, \toolname requires a varying time to verify the synthesized solution, ranging from 0.8 seconds to 15.1 seconds. This highlights the strength of \toolname, being applicable for generating a range of different C programs based on formal and natural language specifications.

Column ``Tot\_Time'' highlights the total time required to generate solutions, which varies significantly across problems. For solved problems, the fastest total runtime is 5 seconds for problem 5, while the longest is 388 seconds for problem 6. This shows that the tool is fast enough to be integrated into a common development process.

\begin{figure}[t]
    \centering
\begin{lstlisting}[language=C, escapeinside={(*@}{@*)}, style=CStyle, basicstyle=\footnotesize\ttfamily, breaklines=true]
(*@\textcolor{blue}{Ground-truth Solution}@*)
void calculateMinimumExamsToResitForGivenSum(int n, int k, int *out)
{
    if (3 * n - k <= 0)
        *out = 0;
    else
        *out = 3 * n - k;
}
(*@    \hrulefill@*)
(*@\textcolor{blue}{Final Synthesized Candidate}@*)
void calculateMinimumExamsToResitForGivenSum(int n, int k, int *out)
{
    *out = (k + 1) / 2;
}
\end{lstlisting}
    \caption{The ground-truth solution and final generated program candidate for problem 12, which does not verify. Given a budget $k$ and $n$ exams, the code minimizes the minimum number of failed exams. See \href{https://codeforces.com/problemset/problem/259/b}{Codeforces} for details.}
    \label{fig:non-verified-code}
\end{figure}

Both unsolved problems (12 and 15) have a significantly higher total time spent. When \toolname fails to generate a verified program, the tool iterates ten times and generates ten candidates for each iteration. This totals to over 100 generated candidates. As the \gls{wp} and \gls{rte} plugins of Frama-C attempt to verify each of the generated program candidates, the tool takes a long time to handle unsolved problems. For problem 12, as shown in \Cref{fig:non-verified-code}, the generated candidate does not meet the formal specification, even after code improvement steps. As per our manual analysis, the \gls{llm} repeatedly includes loops in the synthesized solutions, while the employed prompt mentions that loops are not allowed. The second unsolved problem (problem 15) is solved by employing GPT-4o. This shows that \toolname is able to take advantage of the power of more advanced \glspl{llm} to generate more complex verifying programs; see \Cref{sec:resultsrq3} for more details. 

\begin{tcolorbox}[conclusionbox]
\textbf{Answer to RQ1: \RQone}\\
\toolname proves effective, solving 13 out of 15 problems in \dataset. During the initial code generation step, nine problems are solved, which are improved to 13 throughout the feedback iterations. This demonstrates the capability of \toolname to refine solutions through feedback. Overall, in this experiment, \toolname showcases its potential to automate the synthesis of verified C programs.
\end{tcolorbox}

\subsection{Results for RQ2}
Next, we investigate how different types of input specifications in the prompt affect the ability of \toolname to generate formally verified C programs. Table \ref{tab:specification_types_comparison} presents the results of using a natural language specification, a formal specification, and both specification types when running \toolname. The column ``Specification\_Type`` presents the specification type used when prompting the \gls{llm}. Columns ``Initial`` and ``Improvement'' present the number of problems solved after the initial code generation and improvement steps, respectively. Lastly, column ``Tot\_Time'' presents the total time taken by \toolname to solve all problems with the given specification type.

In the initial code generation step, the natural language prompt solves most problems (10 out of 15), followed by using both specification types (9 out of 15). This shows that using only natural language performs the best when prompting the \gls{llm} without iterative feedback. After the code improvement iterations, using natural language and both types of specifications in the prompt solves 13 problems. This entails that after using the code improvement iterations, including a formal specification or not does not influence the number of solved problems in \dataset. Only using a formal specification performs the worst in both initial code generation and code improvement. Removing the natural language description from the prompt significantly reduces the effectiveness of \toolname. This suggests that natural language specifications are crucial for code generation as \glspl{llm} are primarily trained on such data. While formal specifications alone perform weaker due to limited context, their combination with natural language improves problem-solving efficiency and resource usage, justifying the prompting strategy used by \toolname.

When investigating column ``Tot\_Time'', we see that the configuration using both natural language and formal specifications is the most time-efficient at 3,722 seconds, followed by natural language at 5,447 seconds. This shows that including both specification types reduces the time needed to run \toolname. The reason is that when using natural language only, \toolname spends more time on unsolved problems. For example, this specification type spends 2,892 seconds attempting to solve problem 4. Using only formal specifications takes significantly longer, with a total time of 10,538 seconds. This is because using only formal specifications in the prompt cannot solve seven problems. As we discussed in \Cref{sec:resultsrq1}, if \toolname cannot solve a problem, a lot of time is spent verifying each of the generated candidate solutions, leading to a high time spent per unsolved problem. \toolname demonstrates an effective design by leveraging both specification types, solving the most number of problems in the least amount of time.

\begin{tcolorbox}[conclusionbox]
\textbf{Answer to RQ2: \RQtwo}\\
Experiments show that \toolname performs best when combining natural language and formal specifications, solving most problems in the least time. Natural language specification is important as \glspl{llm} are primarily trained on such data. Formal specifications alone yield weaker performance, but combining both specification types improves efficiency, enabling \toolname to balance problem-solving effectiveness with resource usage.
\end{tcolorbox}
\vspace{-2mm}
\subsection{Results for RQ3}
\label{sec:resultsrq3}

\begin{table*}[t]
\caption{The results of running \toolname with different configurations on \dataset.}
\begin{center}
\begin{tabular}{l|rrrll|rr }
\toprule
\textbf{Configuration\_ID} & \textbf{Candidates} & \textbf{Temperature} & \textbf{Iterations} & \textbf{Prompting Method} & \textbf{LLM} & \textbf{Solved} & \textbf{Tot\_Time}\\ 
\midrule
A (default)   & 10 & 1 & 10 & One-shot & GPT-3.5-turbo & 13 & 3,722 s\\
B  & 1  & 0 & 10 & One-shot & GPT-3.5-turbo & 7 & 2,062 s\\
C  & 100 & 1 & 0 & One-shot & GPT-3.5-turbo & 12 & 5,479 s\\
D    & 10 & 1 & 10 & Zero-shot & GPT-3.5-turbo & 13 & 5,404 s\\
E   & 10 & 1 & 10 & One-shot & GPT-4o & 14 & 830 s\\
F   & 10 & 1 & 10 & One-shot & Llama-3.1-70B & 15 & 3,782 s \\
    \bottomrule
\end{tabular}
\label{tab:configurationscomparison}
\end{center}
\vspace{-8mm}
\end{table*}

Table \ref{tab:configurationscomparison} presents the various configurations in our ablation study that are used to assess the impact of different parameters on the performance of \toolname. Each configuration is identified by a unique identifier presented in column ``Configuration\_ID''. The ``Candidates'' column specifies the number of candidate programs generated per \gls{llm} invocation, and the ``Temperature'' column denotes the used temperature, which controls the randomness in program generation. The ``Iterations'' column lists the maximum number of code improvement iterations, and the ``Prompting Method'' column indicates whether an example of the task is included. Lastly, column ``LLM'' specifies the \gls{llm} used.

The remaining columns present the results. The ``Solved'' column shows the final number of solved problems after running \toolname. The ``Tot\_Time'' column captures the total time spent (in seconds) for code generation and verification per configuration. For example, the default configuration in row A, as presented in \Cref{sec:resultsrq1}, generates ten candidate programs per \gls{llm}-invocation using a temperature of 1. This configuration uses a maximum of ten code improvement iterations and employs a one-shot learning approach. The default configuration solves 13 problems in 3,722 seconds. 

Using configuration B, we evaluate the effect of a lower temperature. Configuration B, with a temperature of 0, solves seven problems after iterations, significantly fewer than configuration A. This demonstrates the benefit of a higher temperature, as it allows \toolname to explore a broader range of program candidates, increasing its effectiveness in generating verifying programs. Configuration B completes in 2,062 seconds, compared to 3,722 seconds for configuration A. This is due to configuration B verifying at most 11 program candidates per problem compared to over 100 in configuration A. This highlights that although configuration A uses more time, the increase in problems solved makes using a higher temperature an effective choice for \toolname.

The effect of removing the one-shot example is analyzed in configuration D. Both configurations A and D solve 13 problems after all iterations. However, configuration A has a lower runtime (3,722 vs 5,403 seconds). Including a one-shot example uses less time when solving problems, helping towards integrating \toolname in a development process. 

Configurations E and F explore the impact of using different \glspl{llm}. Configuration E, which employs GPT-4o, outperforms GPT-3.5-turbo in both effectiveness and time efficiency, solving 14 problems in 830 seconds. Similarly, configuration F uses Llama-3.1-70B, solving all 15 problems with a runtime of 3,782 seconds. These results highlight that \toolname performs well with both open and closed-source \glspl{llm}. This makes \toolname suitable for diverse deployment scenarios.

\begin{tcolorbox}[conclusionbox]
\textbf{Answer to RQ3: \RQthree}\\
The findings indicate that \toolname performs best using a temperature of 1 and generating multiple candidates. Additionally, we find that the iterative approach of \toolname is better than generating more program candidates in the initial code generation step. Moreover, \toolname proves effective with both open- and closed-source \glspl{llm}, solving 14 problems with GPT-4o and all 15 with Llama-3.1-70B, showcasing its versatility with advanced \glspl{llm}.
\end{tcolorbox}

\section{Threats to validity}
\label{sec:threats}

\subsection{Construct Validity}
A threat to the construct validity of our study arises from data leakage and the non-deterministic behavior of \glspl{llm}. Since the problems used in this study are derived from publicly available sources such as Codeforces, it is plausible that the \glspl{llm} may have encountered similar problems or solutions during training. This could lead to an overestimation of their problem-solving capabilities, as they might recall or adapt existing solutions rather than independently generate them. Despite this threat, the data leakage issue in our study is limited as our manually crafted formal specifications were not publicly available before this study. This means the \gls{llm} has not seen these formal specifications in its training dataset.


\subsection{Internal Validity}
The verification process in \toolname depends on the \gls{wp} and \gls{rte} plugins of Frama-C. We run these plugins using the solvers Alt-Ergo, CVC4, and Z3 to leverage and combine their strengths. However, these solvers operate under specific timeouts and step limits. If a solver fails to verify a goal within these constraints, the solution may still be correct, but the solvers cannot confirm it. To mitigate this, we ensure that the ground-truth solution meets the formal specification and thus verifies. This proves that there is at least one solution that can be generated by the \gls{llm} and formally verified by Frama-C. 

\subsection{External Validity}
The limited dataset of 15 problems constrains the generalizability of our findings. Additionally, the problems exclude loops from the analysis. This constraints the applicability of \toolname to more general and complex programming problems. Furthermore, our evaluation is restricted to single-function programs. In the future, expanding the scope to multi-function programs and incorporating more complex data structures would better reflect real-world software development.

The limited dataset of 15 problems constrains the generalizability of our findings. Moreover, the problems exclude loops from the analysis and are not based on safety-critical code, which constrains the applicability of \toolname to more general and complex programming problems.

\section{Related Work}
\label{sec:related-work}

\toolname is the first \glspl{llm}-based tool to automatically generate formally verified C code using an iterative approach. However, other works have explored combining \glspl{llm} with formal methods or employing iterative improvement techniques, which we review in this section. 

\subsection{Formal Specifications-based Code Generation with LLMs}
\label{related-work-codegen}

The closest work to \toolname is \texttt{SynVer} \cite{mukherjee_towards_2024}, a recently published framework for synthesizing and formally verifying C programs. Similar to \toolname, SynVer invokes an \gls{llm} using a natural language specification and a formal specification. The strength of Synver lies in its ability to synthesize complex programs, handling, for example, recursion, which \toolname does not support. SynVer employs a human-in-the-loop approach to support these complex programs, which requires manual intervention to refine and verify programs. In contrast, \toolname prioritizes full automation, iteratively refining candidate programs without human involvement. Additionally, \toolname generates multiple candidate programs and uses test cases to rank the most promising candidate. While both tools share the goal of leveraging \glspl{llm} for program synthesis and verification, they target different levels of automation and complexity.

Patil et al. \cite{patil_towards_2024} propose an iterative \gls{llm}-based framework named \texttt{spec2code} that generates C code from specifications. They do not present any tool and, therefore, manually conduct studies that show promising results. Unlike \texttt{spec2code}, we present a tool that automates verified code generation using \glspl{llm}. Another work investigating the use of formal specifications to generate code is \texttt{SpecEval} \cite{ma_speceval_2024}, which analyzes how well \glspl{llm} understands these specifications. \texttt{SpecEval}, similar to \toolname, generates programs using \glspl{llm}. However, our tool iteratively improves on previously generated program candidates to correct past mistakes.

 Ahrendt et al. \cite{ahrendt_tricotriple_2022} propose a framework named \emph{TriCo} to help users to create code, tests, and specifications simultaneously. Similarly, Sun et al. introduce \emph{Clover} \cite{sun_clover_2024}, which combines \glspl{llm} and formal methods to check consistency between formal specifications, docstring, and code. As these works require formal specifications, several works have investigated automatically generating these based on code \cite{granberry_specify_2024, wen_enchanting_2024} and natural language \cite{cosler_nl2spec_2023, zhai_c2s_2020, giannakopoulou_generation_2020}.

\subsection{Traditional Specification-based code generation}
\label{related-work-llm-formal}

Formal synthesis is a longstanding problem in software engineering aiming to generate programs based on formal specifications \cite{jha_oracle-guided_2010, hsiung_formal_2001, jullig_applying_1993}. Traditional techniques employ deductive synthesis, where programs are derived directly from formal specifications\cite{manna_deductive_1980}. While this method guarantees correctness by construction, it suffers from poor scalability due to the computational complexity of large and complex programs. To address this, Solar-Lezama \cite{solar-lezama_program_2008} introduces sketching, a technique where developers provide partial implementations to guide the synthesis process, reducing synthesis time. Building on this, Alur et al. \cite{alur_syntax-guided_2013} propose \gls{sygus}, which combines syntactic constraints with semantic correctness to improve program generation. 

Over the years, various approaches have been developed to enhance the automatic code generation process, including inductive learning \cite{jha_theory_2016}, oracle-guided synthesis \cite{jha_oracle-guided_2010}, and proof-theoretic synthesis \cite{srivastava_program_2010}. Tools like Fiat \cite{delaware_fiat_2015} refine declarative specifications into functional programs that are correct by construction. Similarly, Li et al. \cite{li_formal_2017} demonstrate the synthesis of verified code from timed automata models, ensuring behavioral correctness while bridging formal models and real-world implementations. Murphy et al. combine these traditional techniques with \glspl{llm} to generate candidates \cite{murphy_combining_2024}.

Contrary to traditional specification-based code generation techniques, \toolname is the first iterative tool to integrate \glspl{llm} for fully automated generation of formally verified C programs. The key difference with traditional methods is that \toolname uses \glspl{llm} to generate the programs and feedback from formal methods to improve faulty generated programs.

\subsection{Iterative Code Improvement with Large Language Models}
\label{related-work-iterative prompting}

Many works employ iterative code improvement as a method for enhancing \gls{llm}-generated programs, leveraging techniques such as automatic program repair \cite{monperrus_automatic_2018, hidvegi_cigar_2024, nilizadeh_exploring_2021}, counterexample-guided synthesis \cite{alur_syntax-guided_2013}, and feedback from compilers, verification tools, or human reviewers \cite{charalambous_new_2023, fakih_llm4plc_2024, xia_keep_2023}. For example,  Jha et al. \cite{jha_dehallucinating_2023} explore providing counterexamples as iterative feedback to mitigate hallucinations of \glspl{llm}. These counterexamples can be derived from formal verification tools \cite{pace_counter-example_2004} or from failed test cases \cite{liu_is_2023}, allowing the \gls{llm} to improve candidate programs based on counter-examples.

Fan et al. highlight that \gls{llm}-generated code frequently suffers from syntax errors, incomplete logic, or incorrect solutions, requiring code improvement through feedback \cite{fan_automated_2023}. The prompt to the \gls{llm} can include previous failed attempts to prevent the \gls{llm} from making the same mistake \cite{xia_keep_2023}. Liventsev et al. propose SEIDR \cite{liventsev_fully_2023}, which iteratively improves program candidates using GPT-assisted summarizations of bugs and failing test cases. A balance between iteratively improving and generating new candidates results in the most improved programs. Tang et al. \cite{tang_code_2024} use Thompson Sampling \cite{thompson_likelihood_1933} to pick what candidate program to repair.

Unlike existing iterative code improvement frameworks, \toolname uniquely integrates formal verifier feedback to iteratively improve on program candidates. In these previous works, feedback for \glspl{llm} relies on using counterexamples, whereas \toolname employs information about unproven goals by Frama-C for this purpose.

\section{Conclusion}
\label{sec:conclusion}

This paper introduced a novel \gls{llm}-based tool \toolname, used for generating formally verified C code. It addresses an initial investigation into automatically generating programs in safety-critical domains. 
\toolname employs a two-step process to generate the programs: 
    \begin{inparaenum}[(1)] 
        \item it leverages \glspl{llm} to generate program candidates based on formal and natural language specifications
        \item it iteratively improves previously generated program candidates through compiler and verifier feedback.
    \end{inparaenum}
Each program candidate is formally verified against the provided formal specification, ensuring that only solutions meeting the specification are accepted. Experiments using \toolname on 15 competitive programming problems demonstrate its effectiveness, solving 13 of them. These results demonstrate the feasibility of \toolname in automating the generation of formally verified C code. As a result, \toolname marks a significant advancement in integrating \glspl{llm} with formal verification, addressing the rigorous correctness requirements of safety-critical software development. \\\\

\noindent\textbf{Acknowledgments}
This work has been partially funded by the Advanced Digitalisation Programme of Sweden's Innovation Agency (VINNOVA) as the FormAI project 2023-02671.
\bibliography{main.bib}
\end{document}